\documentclass[published]{JHEP3}
\JHEP{06(2002)049}

\skip\footins = 1\bigskipamount plus 2pt minus 4pt

\newcommand{\eqn}[1]{(\ref{#1})}

\newcommand{\reals}{\mathbb{R}} 
\def\be{\begin{equation}}
\def\ee{\end{equation}}
\def\beq{\begin{eqnarray}}
\def\eeq{\end{eqnarray}}
\newcommand{\del}{\partial}

\title{Cosmological perturbations and short distance physics from
noncommutative geometry}

\author{Fedele~Lizzi, Gianpiero~Mangano, Gennaro~Miele\\
Dipartimento di Scienze Fisiche, Universit\`{a} di Napoli Federico II\\
and INFN, Sezione di Napoli,\\
Monte S. Angelo, Via Cintia, 80126 Napoli, Italy\\
E-mail: \email{fedele.lizzi@na.infn.it},
\email{gianpiero.mangano@na.infn.it},
\email{gennaro.miele@na.infn.it}}

\author{Marco~Peloso\\
Physikalisches Institut, Universit{\"{a}}t Bonn\\
Nussallee 12, D-53115 Bonn, Germany\\
E-mail: \email{peloso@th.physik.uni-bonn.de}}

\abstract{We investigate the possible effects on the evolution of
perturbations in the inflationary epoch due to short distance
physics. We introduce a suitable non local action for the inflaton
field, suggested by Noncommutative Geometry, and obtained by adopting
a generalized star product on a Friedmann-Robertson-Walker
background. In particular, we study how the presence of a length scale
where spacetime becomes noncommutative affects the gaussianity and
isotropy properties of fluctuations, and the corresponding effects on
the Cosmic Microwave Background spectrum.}

\accepted{June 24, 2002}
\keywords{Non-Commutative Geometry, Cosmology of Theories beyond the SM}
\received{March 19, 2002}

\begin{document}

\section{Introduction: noncommutative spaces and deformed products}

The increasing precision of astronomical data is opening the possibility
that the study of several features of our universe may give hints on yet
unknown properties of fundamental interactions at energy scales well beyond
the domain of standard accelerator experiments. For example, it is
conceivable that some cosmological observables, influenced by the very
early stages of the evolution of the universe, may provide clues on the
nature of gravitational interactions at very small distances. Although it
is standard to assume that the scale at which the properties of spacetime
are modified is close to the exceedingly small Planck length $l_P$, many
models have been recently proposed in which this scale (set presumably by
the fundamental string length) is several order of magnitude
higher~\cite{string1,string2,string3}.

Inflation provides a particularly suitable framework for these
considerations, through its predictions for the primordial perturbations.
These perturbations, responsible for structure formation and for the
temperature anisotropies in the Cosmic Microwave Background (CMB), arise as
quantum fluctuations during the inflationary epoch and are stretched to
cosmological scales by the huge expansion. Hence, they are sensitive to
physics at distances at least as small as the horizon size during
inflation, which in the simplest models is about five orders of magnitude
higher than $l_P$, although they are probably generated at even smaller
scales.

This issue has been studied in the last few years by several
authors~\cite{brand0}--\cite{mbc} (other cosmological and
astrophysical implications of new physics at short distance have also
been discussed. See for example~\cite{acp}--\cite{alex2}). Although
they use different models, most of these analyses share the idea that
the appearance of this new length scale can be modelled in terms of a
transition from an ordinary to a ``quantum'' spacetime, in a way which
is reminiscent of the transition from classical to quantum phase
space. Another possibility is to consider a deformation of the usual
Heisenberg algebra~\cite{kmm,kmm2}. The general finding is that short
distance physics may affect both the gaussian character of the
fluctuations, as well as their isotropy.

It has been recently realized~\cite{CDS,SW} that a consequence of
string theory is that the structure of spacetime becomes
noncommutative. The noncommutativity of the vertex operators, in the
low momentum limit, representing the interaction of strings, and whose
low energy limit gives rise to ordinary spacetime~\cite{LSCMP}, leads
to a spacetime which can be described loosely as an analog of a
quantum phase space, in terms of the algebra generated by noncommuting
coordinates
\be
[x^\mu,x^\nu]=i\Theta^{\mu\nu}(x) \, , \label{xmuxnu}
\ee
with $\Theta^{\mu \nu}$ an antisymmetric tensor. This is a particular kind
of a \emph{Noncommutative Geometry}~\cite{NCBook1,NCBook2,NCBook3,ticos}.
Since the coordinates do not commute, a Heisenberg principle is at work,
and the concept of point is no more meaningful. This is similar to what
happens in the quantization of a classical system. In this case the phase
space loses its meaning as an ordinary manifold, and the observable
quantities corresponding to the coordinates in phase space become
selfadjoint elements of the algebra of operators on an Hilbert space. To
each function on phase space is then associated an operator on an Hilbert
space, and the product of functions is the operator product, which is
noncommutative. An equivalent procedure~\cite{bayen} is to consider a
\emph{deformation} of the algebra of functions on the ordinary
phase space. The functions are the same, but their product is not the usual
pointwise one, but one which reproduces the multiplication of operators.
The technique to study noncommutative spacetime is similar. The main tool
is the definition of the deformed algebra. For a constant $\Theta^{\mu
\nu}$, and in a flat space, the $\star$ product~\cite{Moyal1,Moyal2} is defined
as
follows\footnote{If $\Theta^{\mu \nu}$ is non degenerate, the product in
\eqn{starmin} is an asymptotic expansion of the more general integral
expression $(f \star g)(x) = (1/ \pi)^{n}\int_{\reals^n}\int_{\reals^n}\,
\exp\left(2 i (\Theta^{-1})_{\mu\nu} \left( x^\mu y^\nu + y^\mu
z^\nu + z^\mu x^\nu \right)\right)\,f(y)g(z) \,d^n y \,d^n z$. For
a primer of the various expressions of the product
see~\cite[section~3.5]{ticos}.}
\begin{eqnarray}
f \star g &\equiv& \sum_{k=0}^{\infty} \frac{1}{k!} \, \left(
\frac{i}{2} \right)^k \, \Theta^{\mu_1 \nu_1} \cdots \Theta^{\mu_k
\nu_k} \left( \partial_{\mu_1} \cdots \partial_{\mu_k} f \right)
\left( \partial_{\nu_1} \cdots \partial_{\nu_k} g \right)
\nonumber\\
&=& fg+\frac i2 \Theta^{\mu\nu}\del_\mu f\del_\nu g -\frac 18
\Theta^{\mu_1\nu_1}\Theta^{\mu_2\nu_2}(\del_{\mu_1}\del_{\mu_2}f)
(\del_{\nu_1}\del_{\nu_2}g)+O(\Theta^3) \, .\qquad
\label{starmin}
\end{eqnarray}
At zeroth order in $\Theta^{\mu \nu}$ this expression reduces to the
ordinary (commutative) pointwise product while, at higher order,
derivatives of the fields appear. The definition~\eqn{starmin}
generalizes eq.~\eqn{xmuxnu} in that $x^\mu\star x^\nu-x^\nu\star
x^\mu=i\Theta^{\mu\nu}$. A theory on a noncommutative space of this
kind is obtained by substituting all products between fields in the
action by the $\star$ product.  This implies that the theory becomes
\emph{nonlocal}.

Field theories with $\star$ products have been studied extensively,
from the early work of~\cite{DFR,Filk,VG-B,MSV}. For recent reviews
with extensive references see for
example~\cite{Szabo,DouglasNekrasov,KonechnySchwarz1,KonechnySchwarz2}. Most
of the work has been done in the case of flat spacetime, and for a
constant $\Theta^{\mu \nu}$ with only space-like nonvanishing
components, $\Theta^{0i}=0$ (hereafter we use latin indexes for
spatial components).  The main motivation for this choice is that the
theory would be $local$ in time, and therefore unitary. On the other
hand, in order to preserve Lorentz invariance, the commutator of the
$x^\mu$'s must be a tensor, and its transformation under a boost will
produce nonvanishing space-time components. This problem has been
investigated in~\cite{SeibergSusskindToumbas1,GomisMehen} among
others. The solution to this puzzle lies in the fact that a theory
described by a $\star$ product, as we said, is only an effective
theory of a more fundamental theoretical framework, which only in a
low energy limit gives rise to the relatively simple product
\eqn{starmin}. In string theory the components $\Theta^{ij}$
correspond to a magnetic field, while the $\Theta^{0i}$ to an electric
field. A treatment of scattering of strings in an electric
field~\cite{SeibergSusskindToumbas2,GMMS} shows that in this case
unitarity is restored by the presence of massive open strings modes,
or branes, which cannot be neglected. In light of these results the
unitarity condition can be cast in an invariant form, by requiring
\be
\Theta^{\mu \nu} \Theta_{\mu \nu} > 0\,, \qquad
\epsilon_{\mu \nu \rho \sigma} \Theta^{\mu \nu}
\Theta^{\rho \sigma}=0 \,,
\label{unicond}
\ee
i.e.\ there exists a particular frame where only space-like components
of $\Theta^{\mu \nu}$ are nonvanishing. In all other frames brane
modes should be included to guarantee a unitary behaviour of the
dynamics. Another possibility~\cite{BDFP}, which does not make use of
string theory, is based on the framework developed in~\cite{DFR}. In
the rest of the paper we will take $\Theta^{\mu \nu}$ to have only
space-space nonvanishing components.

The $\star$ product defined in eq.~(\ref{starmin}) defines an
\emph{associative} algebra for a constant $\Theta^{\mu \nu}$, and flat
Minkowski metric. The problem of finding a deformed product, in the
most general case of a Poisson manifold, has been formally solved by
Kontsevich~\cite{Kontsevich}. This solution is however at the level of
formal series in the coordinates. Hence, although of great interest
from a mathematical point of view, it is of limited practical
use. Other noncommutative geometries leading to deformed products have
been discussed in various contexts, for example in a class of deformed
products based on three dimensional Lie algebras~\cite{selene}, but
the list of examples is rather lengthy~\cite{fed1,fed2,fed3}. All of
these products are nonlocal.

In this paper we discuss some effects that a nonlocal theory, based on
noncommutative geometry, might have on the evolution of perturbations
produced during the inflationary epoch. As in~\cite{greene1}, we will
start considering a $\star$ deformed action for a scalar field at the
lowest non trivial order in $\Theta^{\mu \nu}$. However, in the spirit
of general covariance, we first introduce in section~\ref{sec2} a
suitable covariant definition for star product, which reduces
to~(\ref{starmin}) in flat space, and discuss our assumptions on the
antisymmetric tensor $\Theta^{\mu \nu}$. We then consider the simplest
case of a quadratic potential for the inflaton field, though the
formalism can be straightforwardly applied to an arbitrary polynomial
potential. In section~\ref{sec3} we deduce the equation of motion for
field fluctuations, which can be cast in a simple form, analogous to
the usual commutative case, up to a term which breaks spatial
isotropy. Finally we consider how our results may affect the CMB
spectrum, and give our conclusions and outlooks.

\section{Noncommuting variables in the inflation era}\label{sec2}

We assume that the effect of a quantum-like behaviour of spacetime at
short distances can be described via the introduction of the non
trivial commutator~(\ref{xmuxnu}) for comoving coordinates $x^\mu
\equiv(t,{\bf x})$ for a flat Friedmann-Robertson-Walker (FRW)
metric.\footnote{We use unit $\hbar=c=1$.}  There are some natural
requirements which limit the choice of the dimensionful quantity
$\Theta^{\mu \nu}$:

\begin{itemize}
\item[(i)] to satisfy unitarity in the comoving frame we take
$\Theta^{0i}=0$;
\item[(ii)] we require that the uncertainty principle, which follows
from eq.~(\ref{xmuxnu}), affects localization \emph{in physical
coordinates} only at some length scale, and is therefore consistent
with a correspondence principle. The simplest form for $\Theta^{ij}$
which realizes this condition in the comoving frame is the following
\be
\Theta^{ij}=\frac{f^{ij}({\bf x})}{ a^2} \, , 
\label{sol1}
\ee
where $a$ is the scale factor of the universe and its exponent
in~\eqn{sol1} is dictated by the above condition;
\item[(iii)] in a homogenous background metric, it is reasonable to
assume a spatially constant $\Theta^{\mu \nu}$. With no loss of
generality we can therefore choose a frame where the only nonvanishing
space-space component of $\Theta^{\mu \nu}$ is
\be
\Theta^{12}(t)=\frac 1{\Lambda^2 a^2} \,, 
\label{theta12}
\ee
with $\Lambda^{-1}$ the noncommutativity length scale.
\end{itemize}
It is interesting to note that, using the expressions for the non
vanishing Levi Civita connections in a critical FRW spacetime
\be
\Gamma_{0j}^{i}=\frac{\dot{a}}{a} \delta_{j}^{i}\,, \qquad
\Gamma_{ij}^{0}= a \dot{a} \delta_{ij} \,,
\ee
expression~\eqn{theta12} implies
\be
D_\rho\Theta^{12}=0 \label{Dtheta} \,.
\ee
with $D_\rho$ the covariant derivative.

We now construct a \emph{covariant}, \emph{nonlocal} action for a
scalar field $\phi$, inspired by noncommutative geometry, with the
introduction of a deformed $*$ product\footnote{We will reserve the
symbol $\star$ for the flat case product defined in \eqn{starmin},
with a constant $\Theta^{\mu \nu}$.}

For a polynomial potential the \emph{effective} action we will consider
will be of the form
\begin{eqnarray}
S &=& \int d^4 x \sqrt{-g} \left[  \frac{1}{2}  \left(
\partial_\mu \phi \right) * \left( \partial^\mu \phi\right) -
\frac{\lambda}{n!} \, \phi *\cdots * \phi \right] 
\nonumber \\
&\equiv& \int d^4 x \sqrt{-g}  \left[  \frac{1}{2}  \left(
\partial_\mu \phi \right) * \left( \partial^\mu \phi\right) -
\frac{\lambda}{n!} \phi^{*n} \right]  ,
\label{act}
\end{eqnarray}
Since spacetime is now curved, we cannot unambiguously use the $\star$
product defined in \eqn{starmin} since it is not manifestly
covariant. This however immediately suggests that the simplest
generalization is via the substitution of ordinary partial derivatives
in \eqn{starmin} with covariant derivatives. We therefore define a
new, curved, $*$ product
\be
f * g \equiv \sum_{k=0}^{\infty} \frac{1}{k!}  \left( \frac{i}{2}
\right)^k \Theta^{\mu_1 \nu_1} \cdots \Theta^{\mu_k \nu_k} \left(
D_{\mu_1} \cdots D_{\mu_k} f \right) \left( D_{\nu_1} \cdots D_{\nu_k}
g \right) .
\label{star}
\ee

The main problem with the definition \eqn{star} is that it defines a
product, in general, nonassociative. In flat space, covariant
derivatives reduce to commuting ordinary derivatives, and with a
constant $\Theta^{\mu\nu}$, the product $\star$ is associative. In a
curved background, the constancy requirement translates into the
condition that $D_\rho\Theta^{\mu \nu}=0$ for all non vanishing
components which appear in (\ref{star}). Since the
solution~\eqn{theta12} satisfies this requirement, this further
justifies our choice.

However the appearance of the commutators $[D_\mu,D_\nu]$, which in a
curved background vanish only on scalar functions spoils
associativity.  Notice that (thanks to the choice \eqn{theta12}) the
nonassociativity appears only at fourth order in $\Theta^{\mu\nu}$,
while noncommutative effects are already present at order
$\Theta^2$. This enable us to apply a kind of
$\Theta^{\mu\nu}$-perturbative analysis to any analytic potential $V
=\sum_n \lambda_n \,\phi^{*n}$, as well as to evaluate scalar field
$n$ point functions, $\langle\phi(t,{\bf x_1}) *\cdots*\phi(t,{\bf
x_n})\rangle$, provided we only consider the lowest nontrivial
(quadratic) contribution in $\Theta^{\mu\nu}$. In the following we
will spell out the conditions under which this is justified.

To conclude this section we note, in passing, that the algebra of the
vertex operators representing fundamental ground state strings is only
associative in the limit of the string parameter $\alpha'\to
0$~\cite{SW} .  Other nonassociative products (of a different kind)
have appeared in the connections between strings and noncommutative
geometry~\cite{CornalbaSchiappa}.

\section{Nonlocal free scalar field in an expanding universe}\label{sec3}

We consider the action~(\ref{act}) for a quadratic potential
\be
S = \int d^4 x \sqrt{-g} \left[ \frac{1}{2} \left( \partial_\mu \phi
\right) * \left( \partial^\mu \phi\right) - \frac{m^2}{2} \, \phi *
\phi \right] = S_0 + \delta S_K + \delta S_V + {\cal O} \left(
\Theta^4 \right) ,
\label{actq}
\ee
where $\delta S_K$ and $\delta S_V$ represent the lowest non trivial
corrections $({\cal O} (\Theta^2))$ to the kinetic and potential terms
in $S$ respectively. The correction to the mass term can be rewritten
as\footnote{We adopt the notation of~\cite{pea}.}
\begin{eqnarray}
\delta S_V &=& - \frac{m^2}{32} \int d^4 x \sqrt{- g}\, \Theta^{\mu
\nu} \Theta^{\rho \sigma} \left( \left[ D_\mu , D_\nu \right] D_\sigma
\phi \right) D_\rho \phi =
\nonumber\\
&=& \frac{m^2}{32} \int d^4 x \sqrt{-g} \, \Theta^{\mu \nu}
\Theta^{\rho \sigma} R^\tau_{\sigma \mu \nu} \partial_\tau
\phi \partial_\rho \phi \, .
\end{eqnarray}
To obtain this expression, we have used that $D_\mu \Theta^{12}=0$ and
assumed the vanishing of the boundary term
\begin{eqnarray}
\int d^4 x \,\sqrt{-g} \, D_\mu V^\mu &=& \int d^4 x \, \partial_\mu
\left( a^3 V^\mu \right) , 
\nonumber\\ 
V^\mu &\equiv& \Theta^{\mu \nu} \Theta^{\rho \sigma} D_\rho \phi
\left( D_\nu D_\sigma \phi \right) . 
\label{maspar}
\end{eqnarray}
After some algebra and using the expressions for the Riemann tensor in
a FRW universe, eq.~(\ref{maspar}) can be cast in the form
\begin{equation}
\delta S_V = -\frac{m^2}{16} \, \int d^4 x \, a^3 \frac{1}{\Lambda^4}
\left( \frac{\dot{a}}{a} \right)^2 \left( \partial_1 \phi \partial^1
\phi + \partial_2 \phi \partial^2 \phi \right),
\label{delv}
\end{equation}
With an analogous computation, one finds
\begin{eqnarray}
\delta S_K &=& \frac{1}{32} \int d^4 x \sqrt{-g} \, \Theta^{\mu \nu}
\Theta^{\rho \sigma} \left( D_\rho D_\tau \phi \right) \left( \left[
D_\mu , D_\nu \right] D_\sigma D^\tau \phi \right)
\nonumber\\
&=& \frac{1}{16} \int d^4 x \, a^3 \frac{1}{\Lambda^4} \left(
\frac{\dot{a}}{a} \right)^2 \biggl[ \partial_m \partial_0 \phi
\partial^m \partial^0 \phi + \partial_m \partial_i \phi \partial^m
\partial^i \phi -
\nonumber\\ &&  
           \hphantom{\frac{1}{16} \int d^4 x \, a^3 \frac{1}{\Lambda^4} \left(
                     \frac{\dot{a}}{a} \right)^2 \biggl[}\!
- 2 \frac{\dot{a}}{a} \partial^m \phi \partial_0 \partial_m \phi +
\left( \frac{\dot{a}}{a} \right)^2 \partial_m \phi \partial^m \phi
\biggr] \, ,
\label{delk}
\end{eqnarray}
with $m=1,2$.

The two terms $\delta S_K$ and $\delta S_V$ break isotropy.  However,
as long as $S_0$ is the leading contribution, the scalar field $\phi$
can be still assumed to be spatially constant up to small quantum
fluctuations. Note that, for our choice of $\Theta^{\mu \nu}$, the
corrections $\delta S$ do not contribute to the action for the
homogeneous component. For this reason, the background evolves as in
the standard case, if the backreaction of fluctuations can be
neglected. To study the evolution of fluctuations, it is
standard~\cite{mfb} to use conformal time $d \eta = d t / a$, and
decompose
\begin{equation}
\phi = \frac{\chi_0 ( \eta )}{ a} + \int \frac{d^3 {\bf k}}{\left(
2 \, \pi \right)^{3/2} a} \left( {\bf a_k} \, \chi_{\bf k} \left( \eta
\right) {\rm e}^{i {\bf k {\cdot} x}} + {\bf a_k^\dagger} \, \chi_{\bf
k}^* ( \eta ) {\rm e}^{- i {\bf k {\cdot} x}} \right) ,
\end{equation}
with ${\bf a_k},{\bf a_k^\dagger}$ satisfying canonical commutation
relations. The action for fluctuations, up to second order, evaluated
in the vacuum after normal ordering, becomes
\begin{eqnarray}
S &=& \int d \eta \, d^3 {\bf k}  \Biggl[ \vert \chi_{\bf k}' \vert^2 +
\left( \frac{a''}{a} - a^2 m^2 - k^2 \right) \vert \chi_{\bf k} \vert^2+ 
\nonumber\\&&
           \hphantom{\int d \eta \, d^3 {\bf k}  \Biggl[}\!
+\frac{H^2}{8\,\Lambda^4\,a^2}  \biggl( - k_\perp^2 \, \vert
\chi_{\bf k}' \vert^2 + k^2 \, k_\perp^2 \, \vert \chi_{\bf k}
\vert^2 + 
\nonumber\\&&
           \hphantom{\int d \eta \, d^3 {\bf k}  \Biggl[
                     +\frac{H^2}{8\,\Lambda^4\,a^2}  \biggl(}\!
+  \left( 10 \, a^2 H^2 - 6 \, \frac{a''}{a} + a^2 m^2
\right)  k_\perp^2 \, \vert \chi_{\bf k} \vert^2 \biggr) \Biggr] \,,
\label{act2}
\end{eqnarray}
where $k^2 \equiv \sum_i k_i^2$, $k_\perp^2 \equiv k_1^2 + k_2^2$, $H=
\dot{a}/a$, and $'$ denotes derivation with respect to $\eta$.

Since we are interested in the dynamics during the inflationary epoch,
we consider a de Sitter Universe,\footnote{In the model we are
considering, $H$ is actually slowly decreasing, so that the spacetime
is not exactly de Sitter. This leads to a small logarithmic decrease
of the spectrum of fluctuations with increasing momentum~\cite{mfb},
which for the purposes of this paper can be neglected.} setting the
origin of time so that $\eta = - 1 /( a H )$. It is
convenient to redefine
\begin{eqnarray}
\chi_{\bf k} ( \eta) &\equiv& \left( 1 - \epsilon^2 \, k_\perp^2 \,
\eta^2 \right)^{-1/2} y_{\bf k} ( \eta)\,,
\nonumber\\
\epsilon^2 &\equiv& \frac{H^4}{8 \, \Lambda^4} \,,
\label{ridef}
\end{eqnarray}
so that, up to a total derivative, the action~(\ref{act2}) becomes
\begin{eqnarray}
S &=& \int d \eta \, d^3 {\bf k} \left( \vert y_{\bf k}' \vert^2 -
\omega_{\bf k}^2 \, \vert y_{\bf k} \vert^2 \right) , 
\nonumber\\
\omega_{\bf k}^2 &\equiv& k^2 - \frac{2}{\eta^2} + \frac{m^2}{H^2 \,
\eta^2} + \frac{\epsilon^2 \, k_\perp^2}{\left(1 - \epsilon^2
k_\perp^2 \eta^2 \right)^2} ,
\label{act3}
\end{eqnarray}
Notice that eq.~(\ref{act3}) is meaningful only to first order in
$\epsilon^2 \, k_\perp^2 \, \eta^2$, since we have neglected higher
order corrections in $\Theta^{\mu \nu}$ in the
action~(\ref{actq}). Our perturbative analysis is thus valid only for
\begin{equation}
\eta^2 < \eta_i^2 \equiv \frac{8 \, \Lambda^4}{H^4 \, k^2} \,.
\label{range}
\end{equation}
This condition can be easily understood as follows. In a de Sitter
Universe, physical scales $a/k$ grow exponentially, while the horizon
$H^{-1}$ is constant. Any given mode $k$ is then well inside the
horizon in the remote past ($\eta \rightarrow - \infty$) and well
outside in the remote future ($\eta \rightarrow 0$). For our choice of
the time origin, the horizon crossing occurs at $\eta_c = - 1/k$. If
we rewrite eq.~(\ref{range}) as
\be
\frac{a/k}{\Lambda^{-1}} > \left( \frac{H^2}{8\,\Lambda^2}
\right)^{1/2} , 
\label{rerange}
\ee
we see that our analysis can be applied to a given mode $k$ only for
times such that the corresponding physical scale is larger than the
length scale $\epsilon^{1/2} \Lambda^{-1}$. Imposing $\eta_i <
\eta_c$ so that we can follow the dynamical evolution of the mode
$k$ at quadratic order in $\Theta^{\mu\nu}$ well before its horizon
crossing, simply amounts to require that $\Lambda > H$, i.e.\ that
the scale of noncommutativity is smaller than the size of the
horizon. Of course this result also shows that any new effect induced
by the length scale $\Lambda^{-1}$ is not negligible provided this
parameter is not much smaller than the horizon
size,~\cite{brand1,brand2,niemeyer1}.

For $\eta \ll \eta_i$, the equation of motion following from
eq.~(\ref{act3}) becomes\footnote{We neglect the mass term
contribution to $\omega_{\bf k}^2$, see eq.~(\ref{act3}). From a
more detailed analysis (i.e.\ taking into account the time dependence
of $H$), one can show that this term is relevant only at the end of
inflation.}
\begin{equation}
y_{\bf k}'' + \left( - \frac{2}{\eta^2} + k^2 + \epsilon^2 k_\perp^2
\right) y_{\bf k} = 0 \,.
\label{eom}
\end{equation}
In the time interval $\eta_i \ll \eta \ll -1/k$, the action of a
given mode $y_{\bf k}$ is approximately the one of a harmonic
oscillator with the adiabatically changing frequency $\omega_{\bf
k}$. Condition $\eta \ll -1/k$ indicates that the mode is well
inside the horizon, so that the expansion of the Universe is at this
stage negligible. Since the frequency is adiabatically changing, we
adopt the standard procedure (see for example~\cite{mfb}) to start
with the adiabatic vacuum state
\begin{equation}
y_{\bf k} = \frac{1}{\sqrt{2 \, \omega_{\bf k}}}\, {\rm
e}^{-i\,\omega_{\bf k} \, \eta} \, .
\label{adia}
\end{equation}

As well known, the choice of initial conditions is a major problem. In
the standard scenario one assumes that modes emerge at the Planck
scale from a stage where quantum gravity effects cannot be
ignored. Subsequent evolution of these modes is then ruled by known
physics. In our analysis, the threshold scale is given by $\eta=
\eta_i$, i.e.\ for an initial physical scale of the mode of the order
of $H/\Lambda^2$. If during these early times the frequency of any
given mode is real and adiabatically changing ($\omega' \ll
\omega^2$), one expects that fluctuations are always ``kept'' in the
adiabatic vacuum~(\ref{adia}), since this minimizes the energy of the
field (see, for example,~\cite{brand1} for a recent
discussion). Imposing $\eta_i\ll -1/k$ allows us to have an
adiabatically changing frequency for a nonvanishing time interval,
from which the choice~(\ref{adia}) is motivated. Although
eq.~(\ref{act3}) may lead to the conclusion that the adiabaticity
condition breaks at $\eta \simeq \eta_i$, we stress that this equation
is valid for $\eta \gg \eta_i$, so that we cannot exclude on the
basis of the present analysis that the change of frequency is indeed
adiabatic also at earlier times.

We thus start from~(\ref{adia}) and solve the evolution
equation~(\ref{eom}). Expanding also the redefinition~(\ref{ridef}) to
first order in $\epsilon^2 k_\perp^2 \eta^2$, one finds
\begin{eqnarray}
\varphi_{\bf k} &\equiv&  \frac{\chi_{\bf k} }{ a} = \frac{H
\,\eta}{\sqrt{2 \, \alpha \, k}} \, {\rm e}^{-i\alpha k\eta} 
\left( 1 - \frac{1}{2} (\alpha^2-1) \, k^2  \, \eta^2 \right)  \left[ 1
- \frac{i}{\alpha \, k \, \eta} \right] , 
\nonumber\\
\alpha &\equiv& \sqrt{ 1 + \epsilon^2 \sin^2 \vartheta} \,,
\label{solc}
\end{eqnarray} where $k_\perp= k \sin \vartheta$ and $\vartheta$
is the angle between $k$ and the third axis. From this result and
eq.~(\ref{ridef}), the scalar fluctuation have the late time amplitude
\begin{equation}
\vert \varphi_{\bf k} \vert = \frac{H}{\sqrt{2} \,
k^{3/2}}  \left( 1 - \frac{3}{32} \frac{H^4}{\Lambda^4} \sin^2
\vartheta \right) ,
\label{fluct}
\end{equation}
up to correction of higher order in $\epsilon^2$.

The first term in~(\ref{fluct}) is the standard result, valid for
fluctuations of a massless field in a de Sitter universe. The second
is instead a new effect induced by the $\Theta^{\mu \nu}$ depending
terms in the action~(\ref{actq}), and explicitly shows the presence of
a preferred direction associated with the nonvanishing component
$\Theta^{12}$. It is worth noticing that this correction does not
decrease at later times, so that the imprint of noncommutativity is
preserved even after the physical scales of the fluctuations have
grown to much larger sizes than $\Lambda^{-1}$. In the next
section, we will discuss a possible observational implication of this
result.

\section{A possible signature in the cosmic microwave background}\label{sec4}

One of the main successes of inflation is the generation of a nearly flat
spectrum of adiabatic scalar metric perturbations. These primordial
inhomogeneities are the seeds of both structure formation and the
temperature anisotropies observed in the CMB . In this section we discuss
how our findings may affect these perturbations.

We first briefly review the standard case. For an isotropic stress energy
tensor, as it is the case for a minimally coupled scalar field, scalar
metric perturbations can be described by a single gauge invariant variable
$\Phi$~\cite{bardeen}. In longitudinal gauge, the perturbed metric can be
written in terms of $\Phi$ as
\begin{equation}
d s^2 = d t^2 ( 1 + 2 \Phi ) - a^2 d {\bf x}^2 \, ( 1 - 2 \Phi ) \,.
\label{metric}
\end{equation}
For a semiclassical treatment of fluctuations, necessary in order to
find their initial normalization, it is convenient to introduce a
second gauge invariant quantity~\cite{mukh}, which is a superposition
of the scalar metric fluctuation $\Phi$ and the inflaton fluctuation
$\varphi$
\begin{equation}
v = a \left( \varphi + \frac{z}{a} \, \Phi \right) ,\qquad
z \equiv \frac{a \dot{\phi}}{H} \, .
\end{equation}
Using Einstein equations, the action for fluctuations up to second
order can be written in a remarkably simple form for $v_k$, the
Fourier transform of $v$
\begin{equation}
S = \frac{1}{2} \int d \eta \, d^3{\bf k} \left[ \vert v_k'
\vert^2 - k^2 \vert v_k \vert^2 + \frac{z''}{z} \vert v_k \vert^2
\right] .
\label{actv}
\end{equation}
This expression further simplifies during inflation, due to the slow
motion of the inflaton field. For inflation to occur, the $\phi$ field
must be in a region of the potential which satisfies the flatness
conditions
\begin{eqnarray}
\tilde{\varepsilon} &\equiv& \frac{M_P^2}{2} \,
\left(\frac{1}{V}\frac{ \delta V}{\delta \phi} \right)^2 \ll 1 \,,
\nonumber\\
\tilde{\eta} &\equiv& M_P^2 \, \frac{1}{V} \frac{\delta^2 V}{\delta
\phi^2} \ll 1 \,,
\label{slow}
\end{eqnarray}
where $M_P$ is the reduced Planck mass $M_P = 2.4 \cdot
10^{18}$\,GeV. As long as the slow roll conditions~(\ref{slow}) hold,
one finds
\begin{equation}
z' \simeq z \, \frac{a'}{a} \left( 1 - \tilde{\eta} +2 \,
\tilde{\varepsilon} \right) \simeq z \frac{a'}{a} \,.
\end{equation}
The action~(\ref{actv}) thus simplifies to
\begin{equation}
S = \frac{1}{2} \int d \eta \, d^3{\bf k} \left[ \vert v_k' \vert^2 -
k^2 \vert v_k \vert^2 + \frac{a''}{a} \vert v_k \vert^2 \right],
\label{actv2}
\end{equation}
which coincides with the action of a (properly normalized) massless
scalar field, see eq.~(\ref{act2}) with vanishing $\Theta^{\mu \nu}$.

In the following we want to study how the effects discussed in the
previous section may modify this standard picture. Admittedly, these
considerations cannot be completely rigorous, since it is likely that
also the fluctuations of the field $\Theta^{\mu \nu} (t)$ should be
included in the calculation. This would amount in considering the
(unknown) dynamics of $\Theta^{\mu \nu}$.

On the other hand, we stress that, at least in the standard case, the
possibility to describe metric scalar fluctuations as the fluctuations
of a massless minimally coupled scalar field is mainly due to the
slow evolution of the inflaton field, i.e.\ to the slow roll
conditions \eqn{slow}, which (as all background quantities) are left
unchanged by the choice $\Theta^{0i}=0$. We thus may proceed in
analogy with the standard case, and \emph{assume} that
eq.~(\ref{act2}), with $m=0$, describes also the dynamics of the
fluctuations $v_k$. In this way, we follow the common procedure (see
e.g.~\cite{brand0}--\cite{mbc}) to assume that nonstandard physics
modifies the dispersion relation of quantum field excitations, and to
use this relation (in our case the frequency $\omega_{\bf k}^2$ given
in eq.~(\ref{act3})) to compute also the evolution of metric
perturbations. From the analysis of the previous section, we find (see
eq.~(\ref{solc}))
\begin{equation}
v_{\bf k} = - \frac{-i H}{\sqrt{2}}\frac{{\rm e}^{ - i \alpha k
\eta}}{\left( \alpha k \right)^{3/2}} \left( 1 - \frac{1}{2}
(\alpha^2-1) \, k^2 \, \eta^2 \right) \left( 1 + i \alpha k \eta
\right) .
\label{vres}
\end{equation}
{}From $v_{\bf k}$ one can define another gauge invariant quantity
(originally denoted as $\phi_m$ in~\cite{bardeen}, and as ${\cal R}$
in~\cite{lyri}) $ \zeta_{\bf k}\equiv H v_{\bf k} / ( a \, \dot{
\phi})$ which represents the metric scalar perturbation on comoving
hypersufaces, on which fluctuations of the stress energy tensor are at
rest. Except on scales where gravitational collapse has taken place,
this primordial quantity is related to the observed CMB temperature
anisotropies $\delta T/T$ through a linear transfer function ${\cal
T}$.  This function depends upon physics at scales which are much
larger than the scale of noncommutativity $\Lambda^{-1}$, and
therefore it is perfectly legitimate to assume that ${\cal T}$ is
rotationally invariant as in the standard case. Hence, by expanding
the temperature anisotropies in terms of spherical harmonics $Y_{lm}$,
one can write~\cite{lyri}
\begin{eqnarray}
\frac{\delta T}{T} &=& \sum_{l=1}^\infty \sum_{m=-l}^l a_{l m} Y_{l m}\,,
\nonumber\\
a_{l m} &=& \frac{4 \pi i^l}{( 2 \pi )^3} \int d^3 {\bf k}\, {\cal T}
( k , l) \, \zeta_{\bf k} \, Y_{lm} \left( \hat{\bf k} \right) .
\label{alm}
\end{eqnarray}
The first issue to discuss is whether the effects of noncommutativity
may produce temperature fluctuations which deviate from a gaussian
distribution. We start introducing the equal time $n$-point
correlation function for the gauge invariant quantity $v({\bf
x},\eta)$
\be
\Delta({\bf x_1},\dots ,{\bf x_n};\eta)\equiv \langle v({\bf
x_1},\eta) *\cdots * v({\bf x_n},\eta) \rangle \,.
\label{deltan}
\ee
To compute this quantity we therefore need a generalized $*$ product
definition acting on (scalar) functions evaluated at different
spacetime points. In analogy with our discussion of
section~\ref{sec2}, it is easy to see that this can be defined, at
quadratic order in $\Theta^{\mu\nu}$, as follows (see for
example~\cite{Szabo})
\beq
\lefteqn{f_1({\bf x_1},\eta) *\cdots * f_n({\bf x_n},\eta) \equiv }\qquad&&
\label{prod2} \\ 
&\equiv&\Biggl( 1 + \frac{i}{2} \sum_{a<b}\Theta^{\mu \nu} D_\mu^a
D_\nu^b - \frac{1}{8} \sum_{a<b,c<d} \Theta^{\mu \nu}\Theta^{\rho
\sigma} D_\mu^a D_\nu^b D_\rho^c D_\sigma^d \Biggr) f_1({\bf
x_1},\eta) \cdots f_n({\bf x_n},\eta)\,,
\nonumber
\eeq
where $a,b,c,d$ run over $1,\dots ,n$ and, as in section~\ref{sec2},
the only non vanishing component is $\Theta^{12}$, satisfying $D_\rho
\Theta^{12}=0$.  Notice that since we are considering an equal time
product, and $\Theta^{12}$ depends on $\eta$ only, this expression is
defined unambiguously. Moreover, reasoning as in section~\ref{sec2},
it is worth stressing that the product (\ref{prod2}) is associative
when applied to scalar functions, so that the $n$-point
function~(\ref{deltan}) is uniquely defined and can be expressed, via
Wick theorem, in terms of lower order correlation functions.

Using~(\ref{prod2}) and the definition of $\epsilon^2$
in~(\ref{ridef}), it is easy to get
\beq
\Delta({\bf x_1},{\bf x_2};\eta) &=& \int \frac{d^3{\bf k}}{2 \pi^3}\, e^{i
{\bf k} {\cdot} ({\bf x_1-x_2})} |v_{\bf k}|^2 (1- 2 \epsilon^2 k^2 \eta^2)
\label{gauss1}\\
\Delta({\bf x_1},{\bf x_2},{\bf x_3},{\bf x_4};\eta) &=& \Bigr(
\Delta({\bf x_1},{\bf x_2};\eta) \, \Delta({\bf x_3},{\bf x_4};\eta)-
\nonumber \\  &&
            \hphantom{\Bigl(}\!
- 4 \int \frac{d^3{\bf k}}{2 \pi^3}\, \int \frac{d^3{\bf k'}}{2
  \pi^3}\,|v_{\bf k}|^2 |v_{\bf k'}|^2 k^2 k'^2 \epsilon^2 \eta^4\times
\nonumber\\&&
                       \hphantom{\Bigl(- 4 \int \frac{d^3{\bf k}}{2 \pi^3}\, \int}\!
\times\sin^2 \vartheta \sin^2 \vartheta' \sin^2\psi e^{i {\bf k} {\cdot}
  ({\bf x_1-x_2})} e^{i {\bf k'} {\cdot} ({\bf x_3-x_4})} \Bigr)+
\nonumber \\& & 
+\, (2 \leftrightarrow 3) + (2 \leftrightarrow 4)
\label{gauss2}
\eeq where $v_{\bf k}$ is given in~(\ref{vres}) and with, as in the
previous section, $k_\perp \equiv k \sin \vartheta$, $k'_\perp \equiv
k' \sin \vartheta'$, and moreover ${\bf k_\perp} {\cdot} {\bf
k'_\perp} \equiv k_\perp k'_\perp \cos \psi$. Notice that the same
result holds for the correlation functions of the Fourier
anti-transform of $\zeta_{\bf k}$, since the factor $ H a/\dot{\phi}$
depends on $\eta$ only.  Eq.~(\ref{gauss2}) shows that, for late times
$\eta \rightarrow 0$, well after horizon crossing for those scales
which are of cosmological interests today, deviation from gaussianity
due to the noncommutativity scale $\Lambda$ is suppressed by the
factor $(k/a \Lambda)^4$. This is actually a consequence of the fact
that in the comoving frame $\Theta^{12}$ scales as $a^{-2}$, so that
its role at energy scales much smaller that $\Lambda$ is expected to
be negligible.\footnote{This result agrees with what has been obtained
in one of the two scenarios considered in ref.~\cite{greene1}, with
$\Theta^{ij} \sim a^{-2}$. The same conclusion should also hold
\emph{a fortiori} in the other ansatz they consider (a step function
behaviour for $\Theta^{\mu\nu}$) since, at the energy scale relevant
for CMB, $\Theta^{ij}=0$, and their $*$ product reduces to ordinary
pointwise product.} It is important to stress again that deviation
from isotropy, encoded in the parameter $\alpha$, see
eq.~(\ref{vres}), is instead unaffected by the evolution after horizon
crossing. Perturbation amplitudes keep in fact their momentum
distribution at horizon crossing until they become again sub horizon
sized at late times. In the following we neglect in~(\ref{gauss1}) all
contributions which vanish in the limit $\eta \rightarrow 0$.

Since we do not expect any sensible deviation from gaussianity, it
remains true that all information on CMB temperature anisotropy is
encoded in the correlation functions $\langle a_{lm}^* \, a_{l'm'}
\rangle $. In the standard case one gets\footnote{Hereafter we neglect
numerical coefficients which factorize in the final
expression~(\ref{cnoi}).}
\begin{equation}
\langle \zeta_{\bf k}^* \, \zeta_{\bf k'} \rangle \propto \vert
v_k \vert^2 \, \delta^3 \left( {\bf k} - {\bf k'} \right)  ,
\label{zcor}
\end{equation}
and for the coefficients of the CMB power spectrum
\begin{equation}
\langle a_{lm}^* \, a_{l'm'} \rangle = C_0 \left( l \right) \delta_{l
l'} \delta_{m m'} \,.
\label{cstand}
\end{equation}
In this last expression, the uncorrelation for $l \neq l'$ is simply a
consequence of the rotational invariance of $\vert \zeta_{\bf k} \vert
$ in eq.~(\ref{zcor}). In our case, from~(\ref{gauss1}), we see that
eq.~(\ref{zcor}) should be replaced by
\begin{eqnarray}
\lefteqn{\int \frac{d^3 {\bf x_1}}{(2 \pi)^{3/2}} \frac{d^3 {\bf x_2}}{(2
\pi)^{3/2}} e^{i {\bf k} {\cdot} {\bf x_1}} e^{-i {\bf k'} {\cdot} {\bf x_2}}
\langle \zeta({\bf x_1},\eta) * \zeta({\bf x_2},\eta) \rangle \simeq}\qquad\qquad&&
\nonumber \\
&\simeq& \langle \zeta_{\bf k}^* \, \zeta_{\bf k'} \rangle \propto \vert
v_{\bf k} \vert^2 \, \delta^3 \left( {\bf k} - {\bf k'} \right)
\propto k^{-3} \left( 1 - \frac{3}{2} \, \epsilon^2 \sin^2
\vartheta \right),
\label{zcor2}
\end{eqnarray}
with $1/k, 1/k' \gg |\eta| \simeq 0$. Therefore we get
\begin{eqnarray}
\langle a_{lm}^* a_{l'm'} \rangle &\simeq& C_0 ( l ) \Biggl\{
\delta_{l,l'}  \delta_{m,m'} \biggl[ 1 + \frac{3}{2}  \epsilon^2
\biggl( \frac{ ( l+1 )^2 -m^2}{( 2 l +1 ) ( 2 l + 3 )} + \frac{
l^2-m^2} {( 2 l - 1 ) ( 2 l + 1 )} - 1 \biggr) \biggr] -\qquad
\label{cnoi}\\&& 
           \hphantom{C_0 ( l ) \Biggl\{}\!
- \Biggl( \frac{3 \epsilon^2 {\tilde {\cal T}_{l l'}}}{2 ( 4l-3 )}
\sqrt{ \frac{ ( (l-1)^2 -m^2 ) ( l^2-m^2 )}{( 2 l - 1 ) ( 2 l + 1 )}}
 \delta_{l-2,l'}  \delta_{m,m'} - l \leftrightarrow l' \Biggr)
\Biggr\} \,,
\nonumber
\end{eqnarray}
where
\begin{equation}
{\tilde {\cal T}_{l l'}} \equiv \frac{\int_0^\infty \frac{d k}{k}
{\cal T} \left( k , l \right)  {\cal T} \left( k , l'
\right)}{\int_0^\infty \frac{d k}{k} {\cal T}^2 \left( k , l \right)}
\sim 1 \,,
\end{equation}
for scales of cosmological interest $l \geq 100$, and for $l' = l
{\pm} 2$.

The correlation~(\ref{cnoi}) between different multipoles $l$ and $l
{\pm} 2$ is the main effect of the \emph{preferred} direction singled
out by $\Theta^{12}$ and, at least in principle, may be tested in
future CMB experiments like PLANCK. As we already noticed, this effect
is not diluted by the expansion, since it is related to momentum
distribution of fluctuations at horizon crossing. Notice that the
breaking of isotropy leads, in our model, to a quadrupole effect
rather than, as one may have expected, to a dipole term. This is due
to the fact that the first modifications induced by the $*$
product~(\ref{star}) are of quadratic order in $\Theta^{\mu\nu}$.

\section{Conclusions}\label{sec5}

In this work we have discussed how cosmological perturbations may be
affected by a modification of standard physics at short distances
motivated by Noncommutative Geometry. These studies in the framework
of inflationary cosmology provide a unique chance to test physical
laws at very small scales, since the perturbations are generated as
quantum fluctuations on sub horizon scales ($d < ( 10^{13} {\rm\,GeV}
)^{-1}$), and then stretched to cosmological sizes.

To study the evolution of perturbations during inflation, we have
adopted a (nonassociative) covariant generalization of the $\star$
product used in flat spaces, characterized by a length scale
$\Lambda^{-1}$ constant in physical coordinates. This parameter is
related to the vacuum expectation value of the antisymmetric tensor
$\Theta^{\mu \nu}$. At physical distances much greater than
$\Lambda^{-1}$ this product reduces to the usual one, and standard
physics is recovered. The deformed action for fluctuations has been
studied at second order in a $\Theta$ perturbative expansion. The
regime of validity of this analysis is carefully discussed in the
paper, as well as associativity, which is preserved at this level of
approximation.  The main effect on fluctuations turns out to be a
modified dispersion relation, which breaks isotropy.

Finally we have discussed how these effects may be possibly detected
in the CMB anisotropy. While the gaussian character of the
fluctuations is preserved, the particular direction taken by the
nonvanishing $\langle \Theta^{i j} \rangle$ breaks isotropy. Since the
main effect is quadratic in $\Theta^{\mu\nu}$, this leads to a
quadrupole contribution to the CMB spectrum. The size of the
quadrupole term is of order $( H/\Lambda )^4$, where $H$ is the Hubble
parameter during inflation. Thus, it can be visible only provided the
scale $\Lambda^{-1}$ is not much smaller than the size of the horizon
during inflation. Forthcoming satellite experiments, with improved
detection accuracy, as well as a wider sky coverage, may give a chance
for detection of this small quadrupole contribution.

\acknowledgments

M.P. acknowledges Stefan Groot Nibbelink for useful discussions. The
work of F.L.\ is supported in part by the \emph{Progetto di Ricerca di
Interesse Nazionale \emph{SInteSi}}. M.P. is supported by the European
Community's Human Potential Programme under contract
HPRN-CT-2000-00131 Quantum Spacetime, and under contracts
HPRN-CT-2000-00148 and 00152.

\end{document}